\documentclass{mhd}     % you may use here any options admissible 
\usepackage{amsmath}
\usepackage{xcolor}
\usepackage[]{graphicx}
\usepackage{float}
\usepackage{subfigure}
\usepackage{subcaption}
\usepackage{physics}
\usepackage[hidelinks]{hyperref}
\usepackage{algorithm}
\mhdhead{0}{0}{1}	% MHD header (vol., issue, first page)

\title{Extraction of initial dynamics of magnetic micro-convection.} 	%Paper title
\author{M. Klevs, A. Tatuļčenkovs, L.Puķina-Slava, G. Kitenbergs} 	%Author's name 
\institute{MMML lab, Department of Physics, University of Latvia, Jelgavas 3, Riga, LV-1004, Latvia} 

\begin{document}

%----------------------- Head
\maketitle
%\dedication{...}			%if any
\begin{abstract}

Complex fluid flows are important in many real-life problems. For an in-depth understanding, new and more elaborate methods of flow description are necessary. Often experimental and numerical data are accumulated in large quantities however only simple flow properties are extracted. Here we show an approach that allows quantitative extraction of the initial dynamics for magnetic micro-convection. It analyzes the initial time dynamics of the Fourier coefficients of the 2D flow snapshots.
We demonstrate its effectiveness in both numeric and experimental data. In addition, the method is refined to account for noise that can be present in real-world data.
This method can serve as a tool to directly compare mathematical models with experiments. 

\end{abstract}

%----------------------- End of Head     

%----------------------- Body       
\section*{Introduction.}
The subject of fluid mechanics is well known for its complexity.
Recent advances in measurement techniques allow us to collect a large amount of data on the peculiarities of fluid flow, but its interpretation and analysis methods remain a very active research field \cite{annurev1,annurev2}.
It is helpful to explore this with a system that is complex, but well understood, from a theoretical perspective.
Such a case is magnetic microconvection, an instability appearing on a miscible magnetic and non-magnetic fluid interface in a Hele-Shaw cell when exposed to an external magnetic field. Initially, both fluids are separated, but when an external magnetic field is switched on, the fluid interface develops a wavy pattern that evolves into shapes that look like fingers that grow and twist into each other. This instability was first observed in 1980 between two immiscible fluids \cite{a.cebersMagnetostaticInstabilitiesPlane1980} and later the same instability was described for miscible fluids \cite{m.m.maiorovMagneticMicroconvectionDiffusion1983}. Since then, this instability has been studied extensively \cite{erglisMagneticFieldDriven2013, kitenbergsMagneticFieldDriven2015,krakov2021} and the model has also been extended to account for buoyancy effects \cite{kitenbergsGravityEffectsMixing2018a}.
The investigations have been continued both for buoyancy relevant systems \cite{krakov2021,pukina-slavaHowGravityStabilises2023b} and for microfluidics with externally driven fluid flows \cite{KitenbergsMF,krakov_2023}.

The correspondence between the theoretical models and real experiments has been shown by comparing several metrics such as fluid velocities in the "finger" formations \cite{erglisMagneticFieldDriven2013, kitenbergsMagneticFieldDriven2015}, dimensionless values for which the instability starts to form \cite{kitenbergsGravityEffectsMixing2018a}, wavelengths of the "finger" patterns \cite{kitenbergsMagneticFieldDriven2015, kitenbergsGravityEffectsMixing2018a} and the width of the average concentration profile \cite{pukina-slavaHowGravityStabilises2023b}. While these methods work quite well, they do not give a lot of information about the precise dynamics observed in the experimental data.

In this paper, we look at a few methods that can potentially give a lot more information about the initial dynamics of the flow instability. It looks at how perturbations from a perfectly flat interface between the non-magnetic and magnetic fluid grow over time and how the initial growth rate is dependent on the wavenumber of the perturbation. The greatest benefit of this approach is that it provides distinctive characteristics that are directly comparable to analytic predictions \cite{kitenbergsMagneticFieldDriven2015, kitenbergsGravityEffectsMixing2018a} and it contains enough information for it to be a potentially better way to assess how closely the mathematical model fits the observed measurements.

Several variations of the proposed method were compared with each other using numerically simulated data as a reference where different amounts of noise was added to simulate real optical measurements. The best method was then applied to data from real optical measurements as a proof of concept. 

% 1) about history and current state
% 2) method to try
% 3) article structure

% Vectors must by typeset in bold and slanted. You should use for this purpose 
% the new command \verb#\bvec#.
% There are also three new commands \verb#\Div#, \verb#\Rot# and \verb#\Grad# 
% for typsetting vector field operations.

\section{Mathematical model}

\begin{figure}[H]
    \centering
    \includegraphics[width=0.6\linewidth]{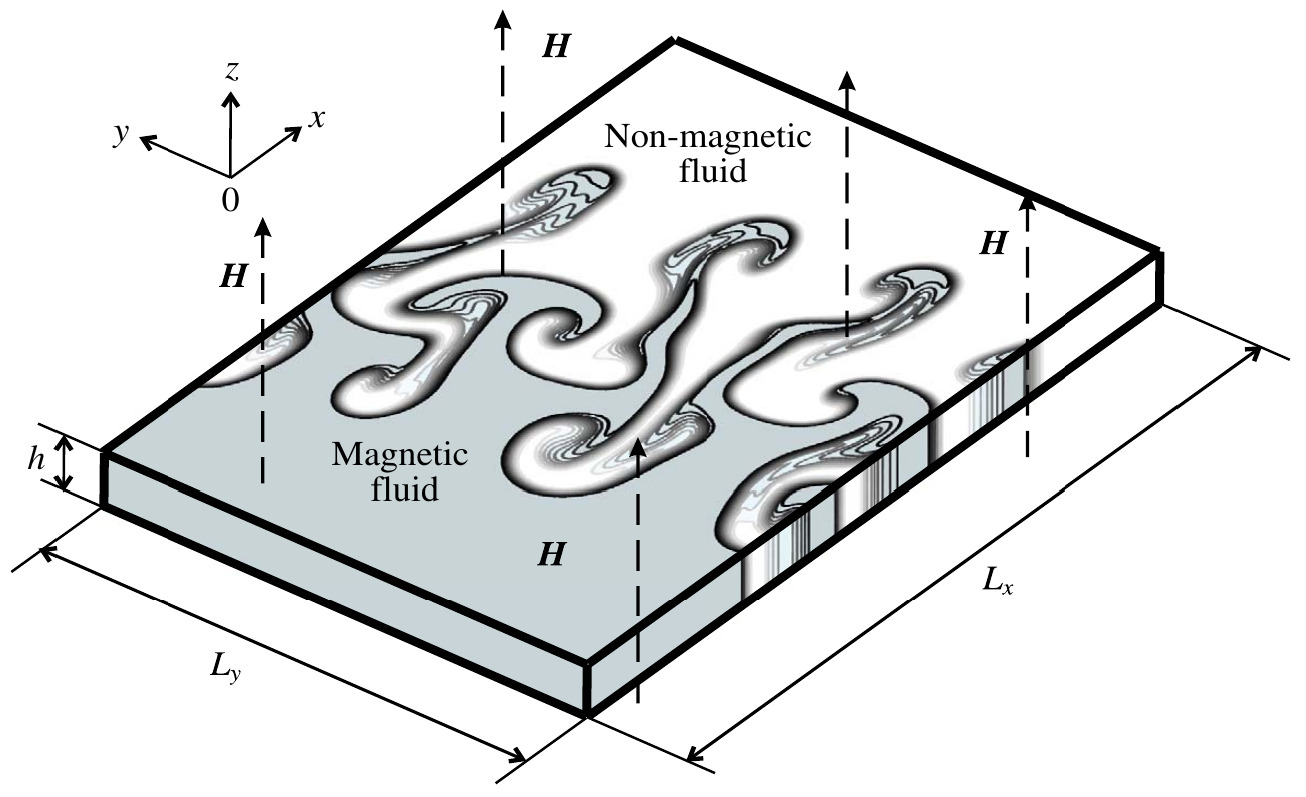}
    \caption{Hele-Schaw cell with an applied external magnetic field \cite{kitenbergsMagneticFieldDriven2015}}
    \label{fig:system_diagram}
\end{figure}

Magnetic microconvection happens in thin sheets where there is a magnetic and non-magnetic fluid as seen in Fig.\ref{fig:system_diagram} where both fluids are separated by a narrow smeared interface, that is created through diffusion.

The magnetic fluid consists of suspended magnetic particles that have intrinsic magnetic dipoles, however due to thermal fluctuations these dipoles are randomly oriented and there is no net magnetic field. 

After a sort period of time a uniform external magnetic field is applied perpendicular to the Hele-Schaw cell which causes the magnetic particles to preferentially orient themselves in the direction of the magnetic field. Assuming that the magnetic particles behave like ideal magnetic dipoles the force acting on a magnetic particle is determined by the gradient of the magnetic field coming from outside of the particle

\begin{equation}
    \bvec{F_m} = \qty(\bvec{m} \cdot \nabla) \bvec{H} 
    \label{eq:mag_dipole_force_discrete}
\end{equation}

where $\bvec{F_m}$ - magnetic force on a single particle, $\bvec{m}$ - magnetic dipole moment, $\bvec{H}$ - magnetic field from the environment. Since the external magnetic field is uniform, the magnetic force comes only from the induced magnetic field coming from the alignment of other magnetic particles. Equation (\ref{eq:mag_dipole_force_discrete}) can also be expressed through force and magnetisation densities in a volume.

It is assumed that the alignment of the magnetic particles is instantaneous compared to any fluid flow therefore the magnetisation field can be taken as a function of the magnetic field. It is expected that the effective magnetic susceptibility of the magnetic fluid is much smaller than unity therefore the model can be further simplified by assuming that the magnetization is proportional to the external magnetic field and the self induced magnetic field doesn't influence magnetisation.

The system is modelled using a 2D Hele-Scaw cell where all values are averaged in depth. The governing equations are as follows \cite{kitenbergsGravityEffectsMixing2018a}.

\begin{equation}
    -\nabla p - \frac{12\eta}{h^2}\bvec{v} - \frac{2M(c)}{h}\nabla\psi_m(c) + \eta\nabla^2\bvec{v} + \Delta\rho c\bvec{g}= 0
    \label{eq:flow_eqn_dimensioned}
\end{equation}

\begin{equation}
    \nabla \cdot \bvec{v} = 0
    \label{eq:div_vel}
\end{equation}

\begin{equation}
    \pdv{c}{t} + \qty(\bvec{v} \cdot \nabla)c = D\nabla^2c
    \label{eq:continuity_c}
\end{equation}

where $p$ - pressure, $\eta$ - kinematic viscosity, $h$ - Hele-Schaw cell depth, $\bvec{v}$ - mean velocity along cell depth, $c$ - concentration field of the magnetic fluid $M$ - magnetisation, $\psi_m$ - magnetic scalar potential, $\Delta\rho$ - density difference between the magnetic an non-magnetic fluid, $\bvec{g}$ - free-fall acceleration from gravity. The magnetic scalar potential is given by \cite{kitenbergsGravityEffectsMixing2018a} where it is expressed in terms of fictitious magnetic surface charges on the upper and lower surfaces of the Hele-Schaw cell and is expressed in a depth-averaged form

\begin{equation}
    \phi_m(\vec{r}) = M_0 \int_S c(\bvec{r}) K(\bvec{r}-\bvec{r'}, h) dS'
    \label{eq:mag_pot_integral}
\end{equation}

where

\begin{equation}
    K(\bvec{r}-\bvec{r'}) = \frac{1}{\sqrt{|\bvec{r}-\bvec{r'}|^2}} - \frac{1}{\sqrt{|\bvec{r}-\bvec{r'}|^2 + h^2}}
    \label{eq:K_mag_pot}
\end{equation}

The governing equations can be re-expressed in a dimensionless form where the characteristic length is $h$, time - $h^2/D$, velocity - $h/D$, pressure - $12\eta D/h^2$, magnetic scalar potential - $M_0h$

\begin{equation}
    -\nabla p - \bvec{v} - 2Ra_m \cdot c\nabla\psi_m(c) + \frac{\nabla^2\bvec{v}}{12} - Ra_g c\bvec{e_y} = 0
    \label{eq:flow_eqn_dimensionless}
\end{equation}

\begin{equation}
    \nabla \cdot \bvec{v} = 0
    \label{eq:div_vel_dimless}
\end{equation}

\begin{equation}
    \pdv{c}{t} + \qty(\bvec{v} \cdot \nabla)c = \nabla^2c
    \label{eq:continuity_c_dimless}
\end{equation}

Here the flow dynamics are determined by two dimensionless numbers. The magnetic Rayleigh number ($Ra_m$) is the ratio between the time scale of the diffusion and the time scale of the magnetic force.

\begin{equation}
    Ra_m=\mu_0M_0^2h^2/12\eta D
    \label{eq:Ram}
\end{equation}

The gravitational Rayleigh number ($Ra_g$) is the ratio between the time scale of the diffusion and the time scale of the gravitational force.

\begin{equation}
    Ra_g=\Delta\rho gh^3/12\eta D
    \label{eq:Rag}
\end{equation}

If the initial interface between both fluids is sharp then there exists an analytical solution to the time evolution of any infinitesimally small perturbations. In the stability analysis done in \cite{kitenbergsGravityEffectsMixing2018a}, the initial perturbation is expressed as a complex exponent that oscillates when going parallel to the initial interface.

\begin{equation}
    \{c,\psi_m,v_x,v_y\}(x,y,t) = \{c_0,\psi_m0,0,0\}(x) + \{c',\psi_m',v_x',v_y'\}(x) e^{iky+\lambda t}
    \label{eq:linear_perturbation}
\end{equation}

For fixed dimensionless numbers $Ra_m$ and $Ra_g$ the exponential growth increment $\lambda$ can be expressed as a function of the spatial wavenumber $k$.

\section{Methods}

\subsection{Numeric model}
% 2D spectral solver without gravity
% vorticity-streamfunction formulation
% The nonlocal magnetic force is calculated in the fourier domain.
% Concentration is initialized with a erf function with a diffusion time t0.
% The interface between both fluids is perturbed by scaled white noise both in the concentration and the streamfunction (velocity)

The numerical model was taken from \cite{kitenbergsGravityEffectsMixing2018a} where the equations (\ref{eq:flow_eqn_dimensionless}), (\ref{eq:div_vel_dimless}), (\ref{eq:continuity_c_dimless}) are solved in the Fourier domain using the vorticity-stream function formulation.

The fluids were initialized to an unmixed state given by the expression

\begin{equation}
    c_0(x,y)=\frac{1}{2}\qty(1-\erf\qty(\frac{x}{2\sqrt{t_0}})) = \frac{1}{2}\qty(1-\erf\qty(\frac{x}{\delta_0}))
    \label{eq:init_c}
\end{equation}

The initialized state includes a small amount of concentration diffusion that is dictated by a dimensionless smearing time $t_0$. The smeared interface is then perturbed by an addition of white noise both for the concentration field and the stream function.

\subsection{Fourier analysis}

While the system behaves linearly multiple perturbations in the form of (\ref{eq:linear_perturbation}) can be summed together without changing the relationship between pairs of $\lambda$ and $k$ 

\begin{equation}
    \{c,\psi_m,v_x,v_y\}(x,y,t) = \{c_0,\psi_{m0},0,0\}(x) + \sum_n\{c'^{n},\psi_m'^{n},v_x'^{n},v_y'^{n}\}(x) e^{ik_ny+\lambda_n t}
    \label{eq:linear_perturbation_sum}
\end{equation}

In practical cases when the measured data is discrete the exponent that is dependent on y can be obtained using the Fast Fourier Transform, however the functions that are dependent on $x$ aren't known. Here it is useful to average the fields over the x direction giving

\begin{equation}
    \frac{1}{L_x}\int_0^{L_x}\{c,\psi_m,v_x,v_y\}(x,y,t) \dd x = \{\overline{c_0},\overline{\psi_{m0}},0,0\} + \sum_n\{\overline{c'^{n}},\overline{\psi_m'^{n}},\overline{v_x'^{n}},\overline{v_y'^{n}}\} e^{ik_ny+\lambda_n t}
    \label{eq:linear_perturbation_sum_profile}
\end{equation}

Another benefit of this transformation is that it reduces any potential noise in the data by averaging over many values. In cases where the captured data isn't periodic over the captured domain it is necessary to use multiply the 1D averaged data with a windowing function before taking the discrete Fourier transform to remove spurious frequencies. In this paper the Blackman windowing function is used. The dynamics of the different wavenumbers can be analysed by looking at how their Fourier coefficients over $y$ change over time. It is sufficient to only look at the concentration field $c$. If the data consists of several snapshots with a time interval $\Delta t$ then $\lambda$ can be calculated by looking at the ratios of the Fourier coefficients between snapshots as follows

\begin{equation}
    \lambda_i = \frac{1}{\Delta t}\ln\left(\frac{1}{N-1}\sum_{n=1}^{N-1}\frac{|a_i^{n+1}|}{|a_i^{n}|}\right)
    \label{eq:fft_growth_increments_from_ratio}
\end{equation}

where $a_i^n$ - i-th Fourier coefficient for the n-th snapshot, $N$ - snapshot count. In an ideal case this can be done with only two snapshots but for noisy data it is better to take an average.

If the data contains additive spatial noise that is constant in time then equation (\ref{eq:fft_growth_increments_from_ratio}) will give inaccurate results because then the ratios of Fourier coefficients would be influenced by a bias both in the numerator and the denominator that doesn't disappear when more timesteps are used in the averaging.

One way to remove constant bias is to look at the time derivative of $a_i$ which has the same exponential term but scaled and the bias will be removed. In practice this can be done by taking the differences between consecutive points in time. The downside of looking at the time differences is that the temporal noise will be significantly amplified.

\subsection{Curve fitting}

Another way to get the growth increments is to preform exponential curve fitting on the Fourier coefficients from (\ref{eq:linear_perturbation_sum_profile}). The exponential curve is defined in the form of

\begin{equation}
    f(t,a,\lambda,b) = a \cdot e^{\lambda \cdot t} + b
    \label{eq:exp_plus_c}
\end{equation}

where $\lambda$ is the predicted growth increment. The curve fitting was done through numeric optimisation where the goal is to minimize the following penalty function

\begin{equation}
    F(a,\lambda,b) = \sum_i \qty(f(t_i,a,\lambda,b) - y_i)^2 + \alpha \lambda^2
    \label{eq:fit_penalty}
\end{equation}

Where $y_i$ is the measured value for a Fourier coefficient in the $i$-th time step and $\alpha$ is a non-negative regularization parameter that biases $\lambda$ towards zero in the cases where an exponent isn't much better at describing the data over a constant value. This regularisation helps with eliminating spurious growth increments when all of the changes in $y_i$ are from noise. If $\alpha$ is chosen to be too large then all predicted values of $\lambda$ will be biased towards zero therefore it is important to choose a value that is small enough to only have a visible effect when the signal is weak.

It is also possible to eliminate the need for bias $b$ by fitting the curve to the time derivative of the Fourier coefficients and setting $b=0$. Similarly to the previous method, taking the derivative significantly amplifies time-dependant noise therefore it needs to be addressed separately.

\subsection{Denoising}

One way to remove noise from the data would be to use an image denoising algorithm for every measured point in time. This will get rid of noise within each separate time value, however there is no guarantee that the measured value within one point is consistent across multiple time steps. Instead the 2D time series data can be denoised as a 3D image to give an even better noise reduction. The changes in the concentration field are expected to be a lot slower than the fluctuations in the noise between each timestep therefore the denoising process won't change the dynamics.

This was done using total variation denoising using split-Bregman optimization \cite{getreuerRudinOsherFatemiTotalVariation2012} where the following expression is minimized

\begin{equation}
    \min_u\sum_{i=0}^{N-1}\qty(\qty|\nabla u_i| + \frac{\beta}{2}(f_i-u_i)^2)
    \label{eq:tv_denoising}
\end{equation}

where $u_i$ - values in the denoised data, $f_i$ - values in the source data, $\beta$ - a positive scalar parameter, $N$ - number of data points. The first term represents the variation and the second term represents similarity to the original image. Decreasing the value of $\beta$ increases the strength of the denoising at the cost to similarity to the original data.

\subsection{Experimental data}

The experimental data was taken from an earlier study. More information can be found in ref.\cite{pukina-slavaHowGravityStabilises2023b}. The setup consist of a small Hele-Schaw cell which is placed inside of an electromagnet coil. The cell is oriented such that the vector of gravity is inside the 2D plane and the slightly more dense magnetic fluid is located in the bottom part of the cell. A light source was located on one side of the Hele-Schaw cell and a camera connected with a microscope was located on the other side.

\subsection{Image processing}

The image time series captures the transmitted intensity of the Hele-Schaw cell, which needs to be converted into an estimate for the concentration field. This is done using the Beer-Lambert law

\begin{equation}
    c(x,y) = \frac{\ln(I_{H2O}) - \ln(I(x,y) - I_{off})}{\ln(I_{H2O}) - \ln(I_{MF} - I_{off})}
\end{equation}

where $I(x,y)$ - measured light intensity, $I_{off}$ - camera bias when there is no light, $I_{H2O}$ - transmitted intensity for pure water, $I_{MF}$ - transmitted intensity for pure magnetic fluid. In this paper $I_{off}$ is assumed to be zero for simplicity. In reality $I_{H2O}$ and $I_{MF}$ also have some dependence on spatial coordinates due to uneven illumination and dust and smudges that can be present on the walls of the Hele-Schaw cell however in this paper they are assumed to be approximately constant due to a lack of calibration measurements for pure water. Small-scale debree was removed using a process described in appendix \ref{appendix:image_despeckling}. The precise background illumination profile was not known therefore it was approximated as constant value. The values of $I_{H2O}$ and $I_{MF}$ were estimated by taking the 98th and 2nd percentiles of the denoised and despeckled input data. Percentiles were used instead of the minimum and maximum intensity values in order to ignore outlier values that can be caused by any remaining noise.

\section{Growth increments from numeric data}

In this work, we examined several methods for extracting the exponential growth increments from a concentration field time series and these methods are discussed in appendix \ref{appendix:method_comparison} where they are labeled as Method 1-8. These methods were compared using data from a numeric simulation for $Ra_m=200$ and $Ra_g=0$ with different kinds of added noise. It is important to look at both time-dependent noise which corresponds to camera sensor noise and time-independent nose (bias) that corresponds with imperfections in the experimental setup such as specs on the Hele-Schaw cell, the interface not being perfectly aligned with the coordinate grid etc. Only the very initial time dynamics were used in the calculations in order for the small perturbation approximation to hold.

In the case where the data is completely noiseless it is possible to accurately extract the growth increments with Method 1 however if noise is added then a more complicated approach is required. The two best performing methods were Method 6 and 8. If the static noise (bias) is small compared to the perturbations then Method 6 gives the best results, however if the static noise (bias) is much larger then Method 8 needs to be used as it is indifferent to static offsets. Both methods use 3D denoising of the 2D time series as a pre-processing step where the parameter $\beta$ determines

\section{Growth increments from experimental data}

Real experimental image sequences were pre-processed using the image intensity scaling and denoising as described above. Methods 1-8 were used on the concentration data and it was determined that Method 8 gave the best results in our specific case. This method gave the best results because the static bias in the Fourier caused by uneven lighting and specks in the Hele-Schaw cell was very large compared to the temporal noise caused by the camera sensor therefore the method's poor performance for the temporal noise wasn't so apparent. The added denoising greatly improved the results of the fitting procedure. The denoising parameter $\beta$ was manually varied and the optimal value $\beta=1$ was determined on a subjective basis by looking for the value that gives the largest growth increments.

Fig.\ref{fig:k_lambda_experimental_comparison_exp_derivative_alpha_0_and_1e-9}a and Fig.\ref{fig:k_lambda_experimental_comparison_exp_derivative_alpha_0_and_1e-9}b show the calculated dimensionless growth increments for fitting penalty parameters $\alpha=0$ and $\alpha=10^{-9}$ from equation \ref{eq:fit_penalty}. Different colored data points represent different experiment runs where the initial dimensionless interface smearing length $\delta_0$ from \ref{eq:init_c} is different. A nonzero fit penalty $\alpha=10^{-9}$ manages to filter out spurious values for high wavenumbers where the signal is dominated by noise.

As $\delta_0$ decreases the peak amplitudes of the growth increments decrease and for large values of $\delta_0$ the growth increments are so small that they are indistinguishable from noise. For $\alpha=0$ the growth increment curves contain a lot of noise 

\begin{figure}[H]
    \centering
    \includegraphics[width=0.75\linewidth]{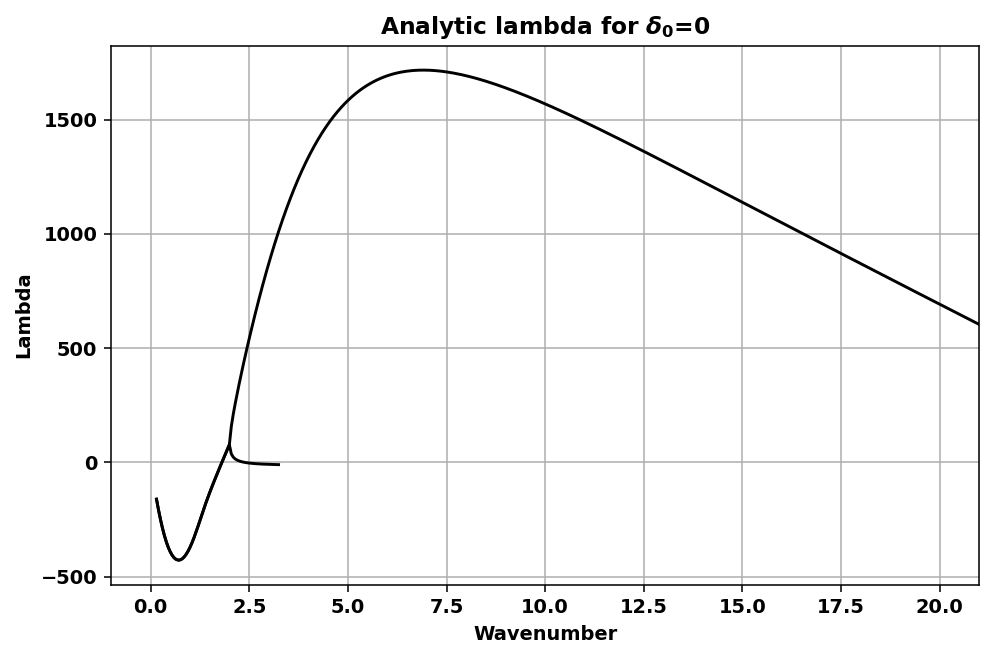}
    \caption{Growth increments given by the theoretic model in \cite{kitenbergsGravityEffectsMixing2018a} using equation (\ref{eq:analytic_k_lambda}) for $Ra_m=1027$ and $Ra_g=2327$.}
\label{fig:k_lambda_theoretical_delta_0}
\end{figure}

It is expected that when $\delta_0$ goes to zero the growth increments should approach the values given by the analytic model in Fig.\ref{fig:k_lambda_theoretical_delta_0}. It is immediately apparent that the theoretical peak growth increment is approximately three times larger than the observed value and that the values for larger wavenumbers decay a lot slower. The much smaller amplitude in the experimental data can possibly be caused by the nonzero $\delta_0$ because in Fig.\ref{fig:k_lambda_experimental_comparison_exp_derivative_alpha_0_and_1e-9}b it can be seen that for small $\delta_0$ the differences in the growth increments from changing $\delta_0$ become large so it is possible that the jump between $\delta_0=0.06$ and $\delta_0=0$ explains the discrepancies. It is also possible that there are some out of plane effects in the fluid flow and in the induced magnetic forces that cause this deviation.

One thing that is quite similar between the calculated and the theoretical values is the wavenumber where the growth increments first become positive which is around $k=2-2.5$. For $\delta_0$ we can see that wavenumbers that are smaller than $k=2.5$ follow a downwards curve that is similar to the one predicted by the theoretical model in Fig.\ref{fig:k_lambda_theoretical_delta_0}. By examining the image sequences it was observed that this decay corresponds to an initial straightening of the interface.

\section{Conclusions}

\begin{figure}[H]
    \centering
    \includegraphics[width=0.49\linewidth]{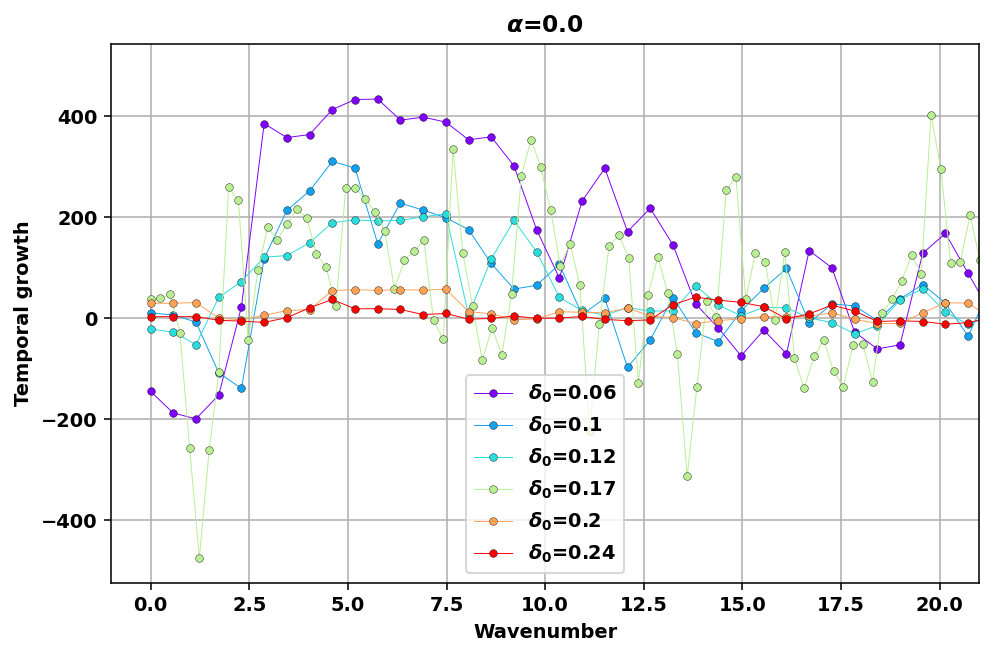}
    % \caption{Calculated growth increments for experimental data with the regularisation parameter $\alpha=0$ where $Ra_m=1027$ and $Ra_g=2327$. Different colors represent different initial interface thicknesses.}
    % \label{fig:k_lambda_experimental_comparison_exp_derivative_alpha_0}
    \hfill
    \includegraphics[width=0.49\linewidth]{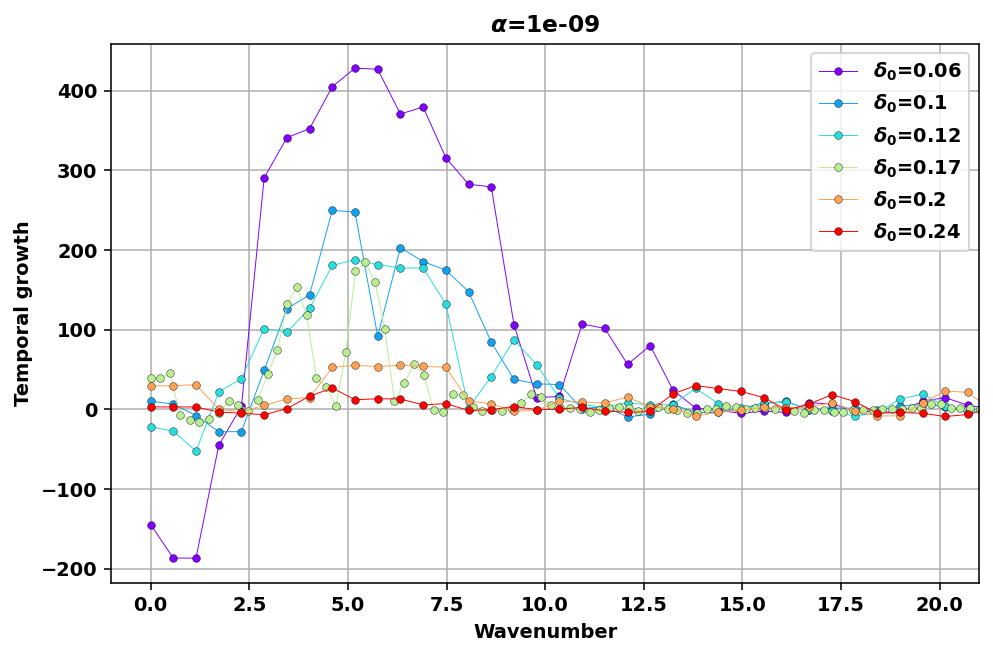}
    % \caption{Calculated growth increments for experimental data with the regularisation parameter $\alpha=10^{-9}$ where $Ra_m=1027$ and $Ra_g=2327$. Different colors represent different initial interface thicknesses.}

    \caption{Calculated growth increments for experimental data where $Ra_m=1027$ and $Ra_g=2327$. Different colors represent different initial interface thicknesses. a) with the regularisation parameter $\alpha=0$, b) with the regularisation parameter $\alpha=10^{-9}$}
    \label{fig:k_lambda_experimental_comparison_exp_derivative_alpha_0_and_1e-9}
\end{figure}

In this study we have shown that it is possible to numerically extract the initial exponential growth increments for different perturbation wavenumbers using a time series of the magnetic fluid concentration field. 
The thickness of the initial fluid interface greatly influences the observed growth increments in Fig.\ref{fig:k_lambda_experimental_comparison_exp_derivative_alpha_0_and_1e-9} where a lower initial thickness $\delta_0$ leads to larger growth and if $\delta_0$ is too large then growth is either very small or impossible to measure with the current noise level. The measured growth increments were much lower than the values given by the theoretical model however this can potentially be explained by the initial interface having a nonzero thickness $\delta_0$ as opposed to the discrete jump that is assumed in the model. The rough similarities in the shape of the growth increment curve for small $\delta_0$ and the theoretical curve gives some confidence in the growth increment extraction method.

While here the results that were gained from experimental data are limited by both camera noise and imperfections in the experiment this study demonstrates that under ideal conditions it is possible to closely reconstruct the growth increment curve. If the concentration field were to be captured more accurately, this method would give more accurate results that can then be further compared with current theoretical predictions. An improved experimental setup would utilize a high bit-depth camera with a higher frame rate. The lighting intensity and camera settings should be adjusted such that the differences in the transmitted intensity between the magnetic fluid and water span most of the camera bit depth. The experimental setup would need to be calibrated by measuring the time-average intensity field with back illumination without the measured sample and also for the back illumination being turned off. Ideally the walls of the Hele-Shaw cell would need to be as clean as possible in order to avoid specks and smudges.

The topic of fluid instabilities remains an active research topic, as multiple recent examples show \cite{Miranda2023,Nagel2024,Mishra_2024}. 
We encourage other researchers to apply this method to analyze their finding in a more descriptive manner.

\Thanks{This research is funded by the Latvian Council of Science, Latvia, projects BIMs, project No. lzp-2020/1-0149, and A4Mswim, project No.lzp-2021/1-0470. }

%-------------------- End of Body

%-------------------- References     

%\printbibliography
\bibliography{bibliography.bib}

\begin{thebibliography}{10}

\bibitem{annurev1}
{\sc S.~L. Brunton, B.~R. Noack, and P.~Koumoutsakos}.
\newblock Machine learning for fluid mechanics.
\newblock {\it Annual Review of Fluid Mechanics\/}, vol.~52 (2020), no. Volume 52, 2020, pp.~477--508.

\bibitem{annurev2}
{\sc L.~Magri, P.~J. Schmid, and J.~P. Moeck}.
\newblock Linear flow analysis inspired by mathematical methods from quantum mechanics.
\newblock {\it Annual Review of Fluid Mechanics\/}, vol.~55 (2023), pp.~541--574.

\bibitem{a.cebersMagnetostaticInstabilitiesPlane1980}
{\sc {A.Cebers} and {M.M.Maiorov}}.
\newblock Magnetostatic instabilities in plane layers of magnetizable fluids.
\newblock {\it Magnetohydrodynamics\/}, vol.~16 (1980), no.~1, pp.~27--35.

\bibitem{m.m.maiorovMagneticMicroconvectionDiffusion1983}
{\sc {M. M. Maiorov} and {A. O. Tsebers}}.
\newblock Magnetic microconvection on the diffusion front of ferroparticles.
\newblock {\it Magnetohydrodynamics\/}, vol.~19 (1983), no.~4, pp.~376--380.

\bibitem{erglisMagneticFieldDriven2013}
{\sc K.~{\=E}rglis, et~al.}
\newblock Magnetic field driven micro-convection in the {{Hele-Shaw}} cell.
\newblock {\it Journal of Fluid Mechanics\/}, vol.~714 (2013), pp.~612--633.

\bibitem{kitenbergsMagneticFieldDriven2015}
{\sc G.~Kitenbergs, et~al.}
\newblock Magnetic field driven micro-convection in the {{Hele-Shaw}} cell: The {{Brinkman}} model and its comparison with experiment.
\newblock {\it Journal of Fluid Mechanics\/}, vol.~774 (2015), pp.~170--191.

\bibitem{krakov2021}
{\sc M.~S. Krakov, A.~R. Zakinyan, and A.~A. Zakinyan}.
\newblock Instability of the miscible magnetic/non-magnetic fluid interface.
\newblock {\it Journal of Fluid Mechanics\/}, vol.~913 (2021), p.~A30.

\bibitem{kitenbergsGravityEffectsMixing2018a}
{\sc G.~Kitenbergs, A.~Tatu{\c l}{\v c}enkovs, L.~Pu{\c k}ina, and A.~C{\=e}bers}.
\newblock Gravity effects on mixing with magnetic micro-convection in microfluidics.
\newblock {\it Eur. Phys. J. E\/}, vol.~41 (2018), no.~11, p.~138.

\bibitem{pukina-slavaHowGravityStabilises2023b}
{\sc L.~{Pu{\c k}ina-Slava}, A.~Tatu{\c l}{\v c}enkovs, A.~C{\=e}bers, and G.~Kitenbergs}.
\newblock How gravity stabilises instability: The case of magnetic micro-convection.
\newblock  (2023), no. arXiv:2310.15323.

\bibitem{KitenbergsMF}
{\sc G.~Kitenbergs and A.~Cēbers}.
\newblock Rivalry of diffusion, external field and gravity in micro-convection of magnetic colloids.
\newblock {\it Journal of Magnetism and Magnetic Materials\/}, vol.~498 (2020), p.~166247.

\bibitem{krakov_2023}
{\sc M.~S. Krakov}.
\newblock Waves and instability at the interface of two flows of miscible magnetic and non-magnetic fluids.
\newblock {\it Journal of Fluid Mechanics\/}, vol.~970 (2023), p.~A11.

\bibitem{getreuerRudinOsherFatemiTotalVariation2012}
{\sc P.~Getreuer}.
\newblock Rudin-{{Osher-Fatemi Total Variation Denoising}} using {{Split Bregman}}.
\newblock {\it Image Processing On Line\/}, vol.~2 (2012), pp.~74--95.

\bibitem{Miranda2023}
{\sc I.~M. Coutinho and J.~A. Miranda}.
\newblock Role of interfacial rheology on fingering instabilities in lifting hele-shaw flows.
\newblock {\it Phys. Rev. E\/}, vol.~108 (2023), p.~025104.

\bibitem{Nagel2024}
{\sc S.~D. Gowen, T.~E. Videb\ae{}k, and S.~R. Nagel}.
\newblock Measurement of pressure gradients near the interface in the viscous fingering instability.
\newblock {\it Phys. Rev. Fluids\/}, vol.~9 (2024), p.~053902.

\bibitem{Mishra_2024}
{\sc A.~Patmonoaji, Y.~Nagatsu, and M.~Mishra}.
\newblock Instability dynamics of viscous fingering interaction on dual displacement fronts.
\newblock {\it Journal of Fluid Mechanics\/}, vol.~995 (2024), p.~A5.

\bibitem{vincentGrayscaleAreaOpenings1993}
{\sc L.~Vincent}.
\newblock Grayscale area openings and closings, their efficient implementation and applications.
\newblock In {\it First {{Workshop}} on {{Mathematical Morphology}} and Its {{Applications}} to {{Signal Processing}}\/} (Citeseer, 1993), pp. 22--27.

\bibitem{damelinSurfaceCompletionImage2018}
{\sc S.~B. Damelin and N.~S. Hoang}.
\newblock On {{Surface Completion}} and {{Image Inpainting}} by {{Biharmonic Functions}}: {{Numerical Aspects}}.
\newblock vol.~2018 (2018), no.~1, p.~3950312.

\end{thebibliography}
\bibliographystyle{mhd}
\newpage
% -------------------- End of References
\lastpageno	% This command sends the number of the last page to 
		% the MHD headline. Please latex your file twice if
		% you have used it.

%%%%%%%%%%%%%  
%	Place your tables and figures here
%%%%%%%%%%%%%

\begin{figure}[H]
    \centering
    \includegraphics[width=0.95\linewidth]{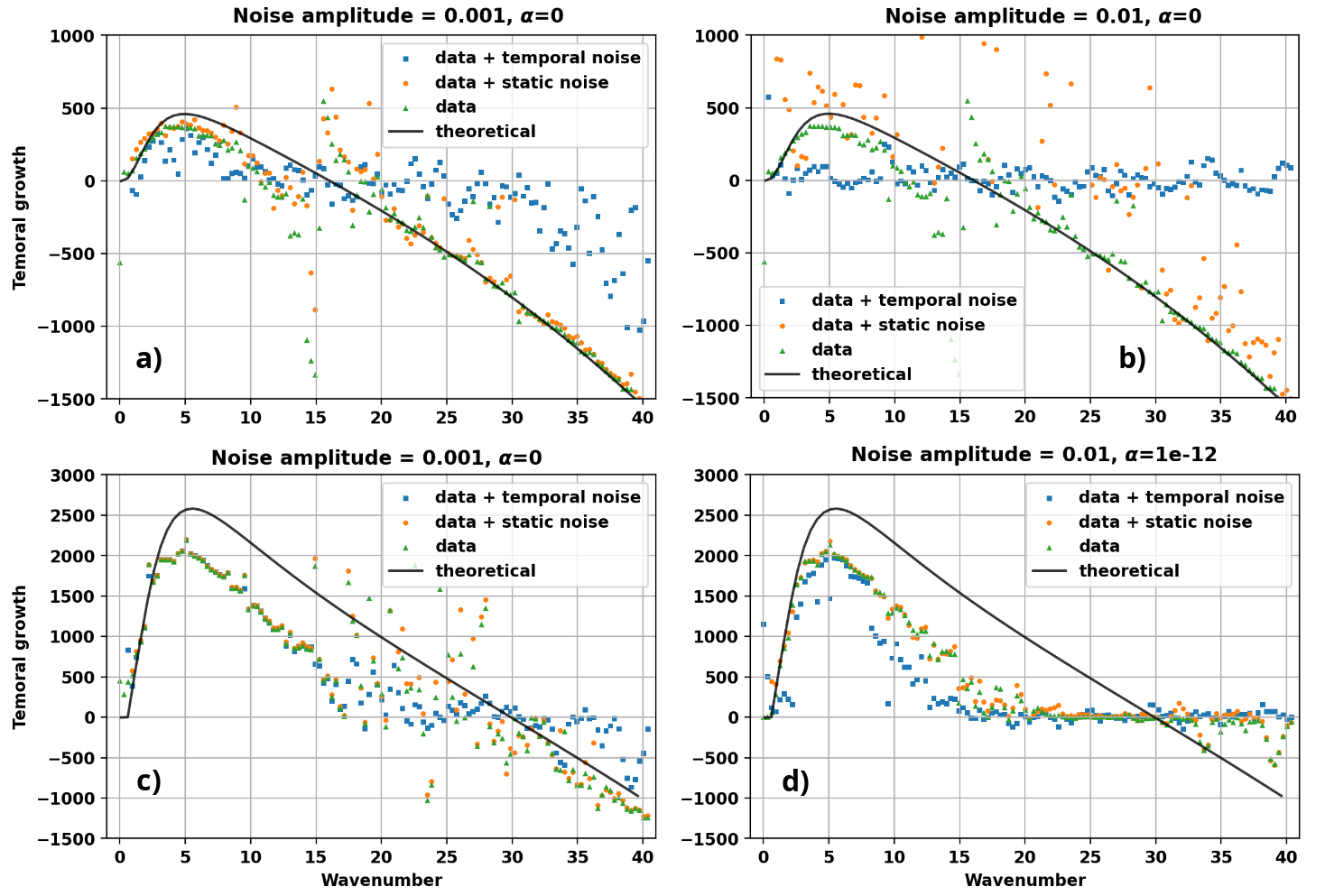}
    \caption{The growth increment dependence on wavelength for different $Ra_m$ under different levels of noise. a) $Ra_m=200$ with the noise amplitude of 0.001 b) $Ra_m=200$ with the noise amplitude of 0.01 c) $Ra_m=200$ with the noise amplitude of 0.001 d) $Ra_m=200$ with the noise amplitude of 0.01}
    \label{fig:combined_figure_200_1e-3_200_1e-2_1000_1e-2_1000_1e-3}
\end{figure}

\newpage

\appendix
\section{Analytic values for the growth increments}
\label{appendix:analytic_increments}

It has been shown in \cite{kitenbergsGravityEffectsMixing2018a} that the relationship between $k$ and $\lambda$ for a linear perturbation in the form of (\ref{eq:linear_perturbation}) assuming a perfectly sharp initial boundary between both fluids can be found using the following expression 

\begin{equation}
\begin{aligned}
s k+R a_m\left(k\left[\frac{g(k(s+m), \infty)}{m}-g(k(s+1), \infty)\right]+\right. \\
\left.+2 \frac{m-1}{m} J(s, k)-\frac{R a_g}{2}\left[\frac{1}{m(s+m)}-\frac{1}{s+1}\right]\right)=0
\end{aligned}
\label{eq:analytic_k_lambda}
\end{equation}

where the parameters $s$, $m$, $J$, $g$ are defined as follows

\begin{gather}
    s = \sqrt{1+\lambda/k^2} \\
    m = \sqrt{1+12/k^2} \\
    J(p,q) = \int_0^\infty e^{-pz}(K_0(z) - K_0(\sqrt{z^2+q^2})) \dd z\\
    g(a,z) = \int_0^z e^{-a\zeta}\ln(1+\zeta^{-2}) \dd\zeta
\end{gather}

Here $K_0$ is the modified Bessel function of the second kind. A specific value of $\lambda$ can be found by fixing $k$, $Ra_m$ and $Ra_g$ and numerically finding the root of the left-hand side of (\ref{eq:analytic_k_lambda}) by varying $\lambda$.

\newpage

\section{Comparison between different growth increment extraction methods}
\label{appendix:method_comparison}

% Lx:             20
% Ly:             20
% Nx:             512
% Ny:             512
% n_snapshots:    200
% dt:             0.00001
% t1:             0.002

There are many ways to approach the extraction of the growth increments and each way responds differently to noise in the input data. It is useful to compare the different methods on data taken from numeric simulations with varying degrees of added noise.

The examined methods were:

\begin{enumerate}
    \item 2D concentration field + no noise, time-dependent noise, static noise $\rightarrow$ DFT coefficients using (\ref{eq:linear_perturbation_sum_profile}) $\rightarrow$ method of ratios (\ref{eq:fft_growth_increments_from_ratio})
    \item 2D concentration field + no noise, time-dependent noise, static noise $\rightarrow$ denoising using (\ref{eq:tv_denoising}) $\rightarrow$ DFT coefficients using (\ref{eq:linear_perturbation_sum_profile}) $\rightarrow$ method of ratios (\ref{eq:fft_growth_increments_from_ratio})
    \item 2D concentration field + no noise, time-dependent noise, static noise $\rightarrow$ denoising using (\ref{eq:tv_denoising}) $\rightarrow$ DFT coefficients using (\ref{eq:linear_perturbation_sum_profile}) $\rightarrow$ time derivative $\rightarrow$ method of ratios (\ref{eq:fft_growth_increments_from_ratio})
    \item 2D concentration field + no noise, time-dependent noise, static noise $\rightarrow$ DFT coefficients using (\ref{eq:linear_perturbation_sum_profile}) $\rightarrow$ exponential fit with bias
    \item 2D concentration field + no noise, time-dependent noise, static noise $\rightarrow$ DFT coefficients using (\ref{eq:linear_perturbation_sum_profile}) $\rightarrow$ exponential fit without bias
    \item 2D concentration field + no noise, time-dependent noise, static noise $\rightarrow$ denoising using (\ref{eq:tv_denoising}) $\rightarrow$ DFT coefficients using (\ref{eq:linear_perturbation_sum_profile}) $\rightarrow$ exponential fit without bias
    \item 2D concentration field + no noise, time-dependent noise, static noise $\rightarrow$ DFT coefficients using (\ref{eq:linear_perturbation_sum_profile}) $\rightarrow$ time derivative $\rightarrow$ exponential fit without bias
    \item 2D concentration field + no noise, time-dependent noise, static noise $\rightarrow$ denoising using (\ref{eq:tv_denoising}) $\rightarrow$ DFT coefficients using (\ref{eq:linear_perturbation_sum_profile}) $\rightarrow$ time derivative $\rightarrow$ exponential fit without bias
\end{enumerate}

\begin{figure}[H]
    \centering
    \includegraphics[width=0.49\linewidth]{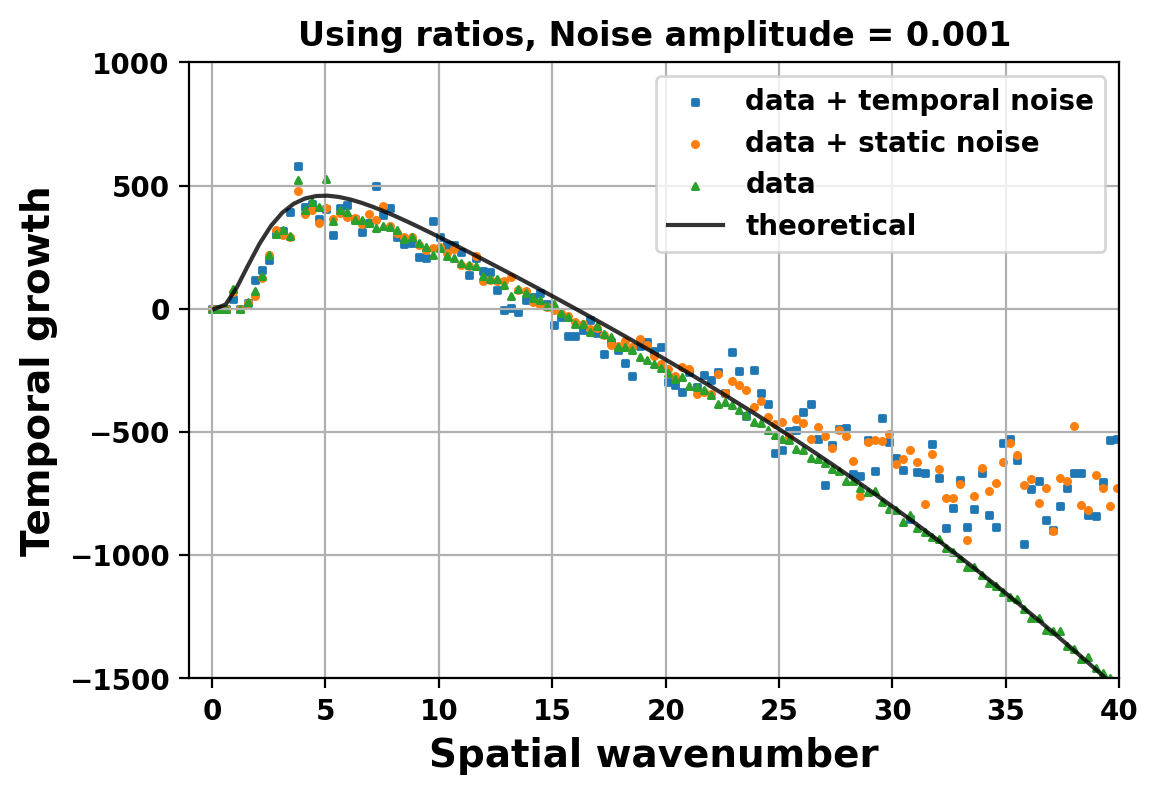}\hfill
    \includegraphics[width=0.49\linewidth]{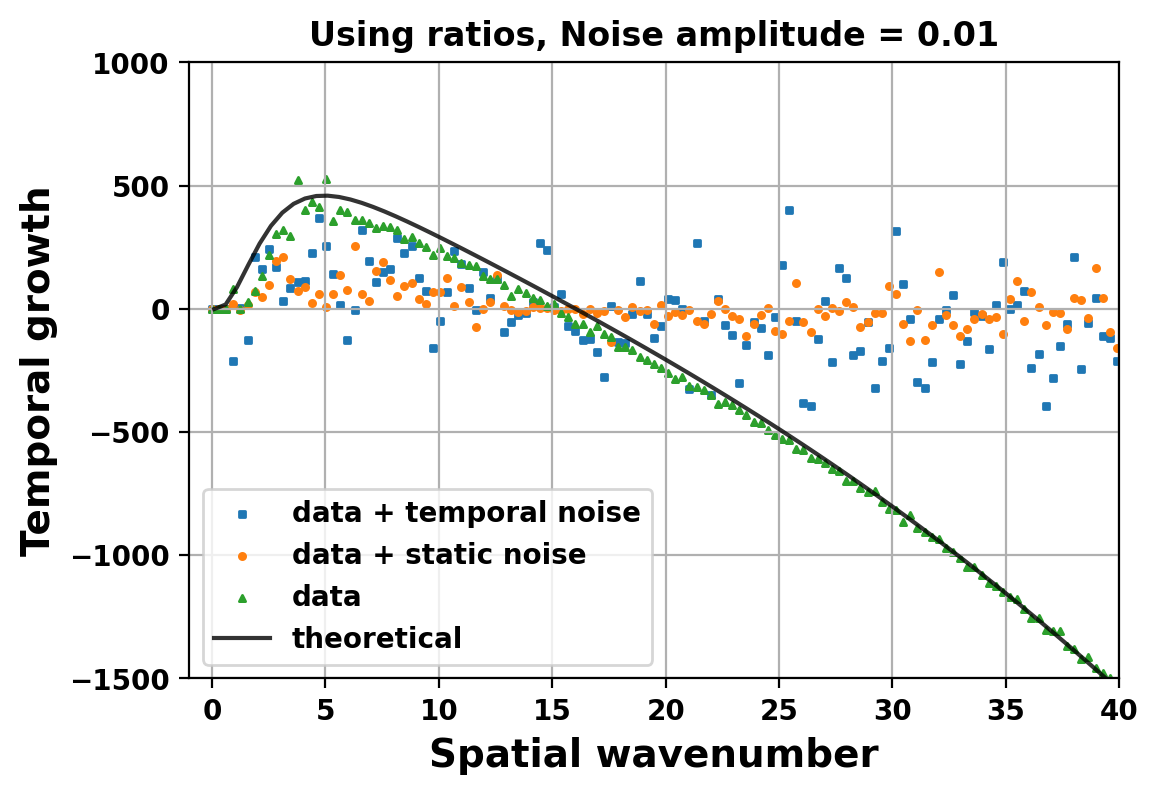}
    
    \includegraphics[width=0.49\linewidth]{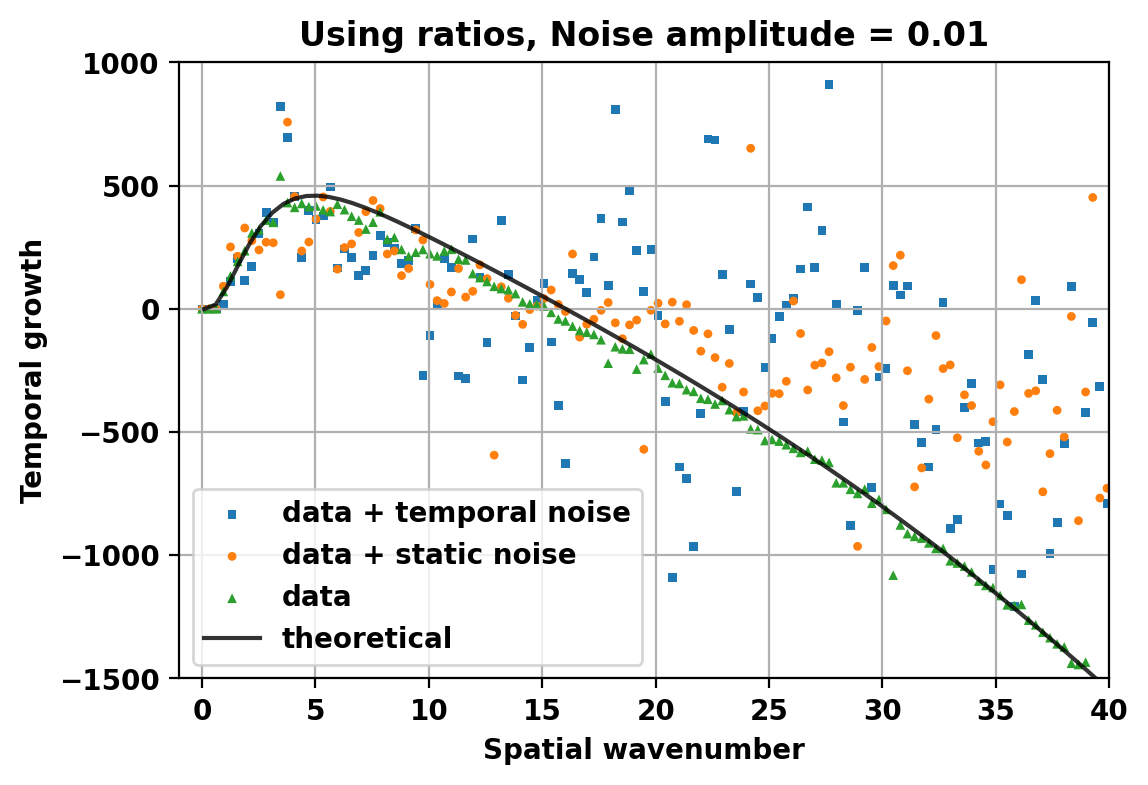}\hfill
    \includegraphics[width=0.49\linewidth]{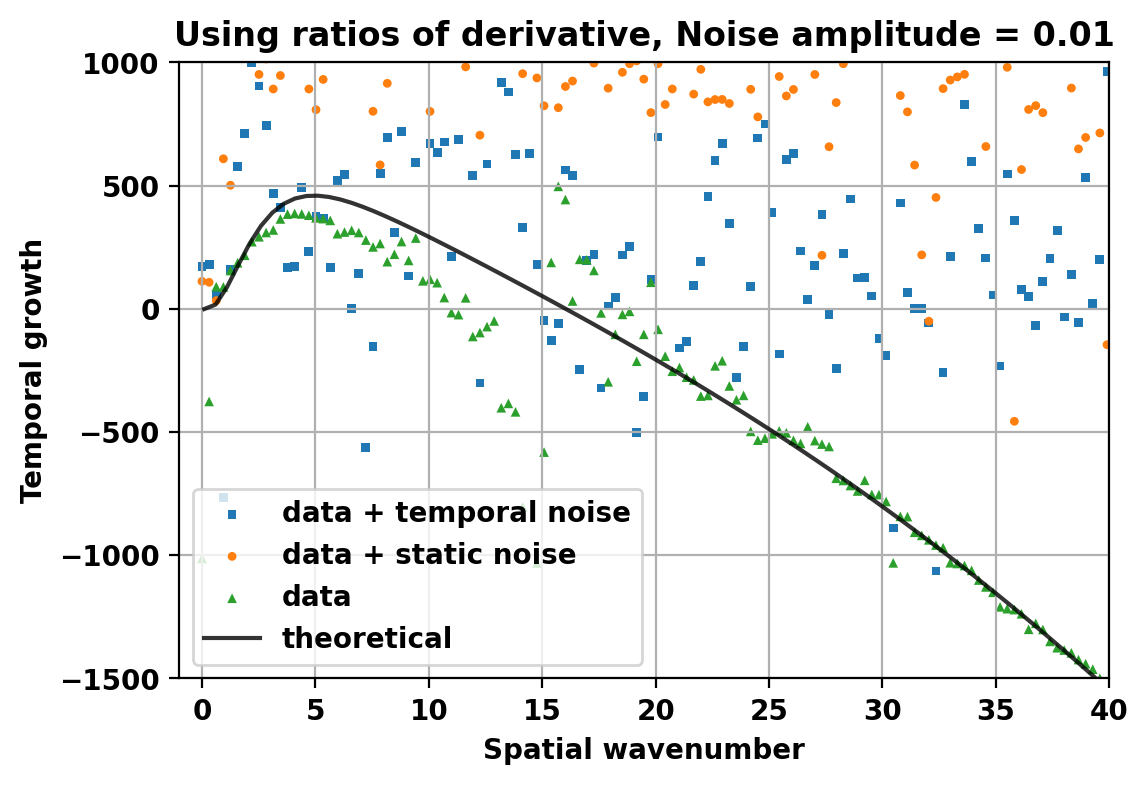}
    
    \includegraphics[width=0.49\linewidth]{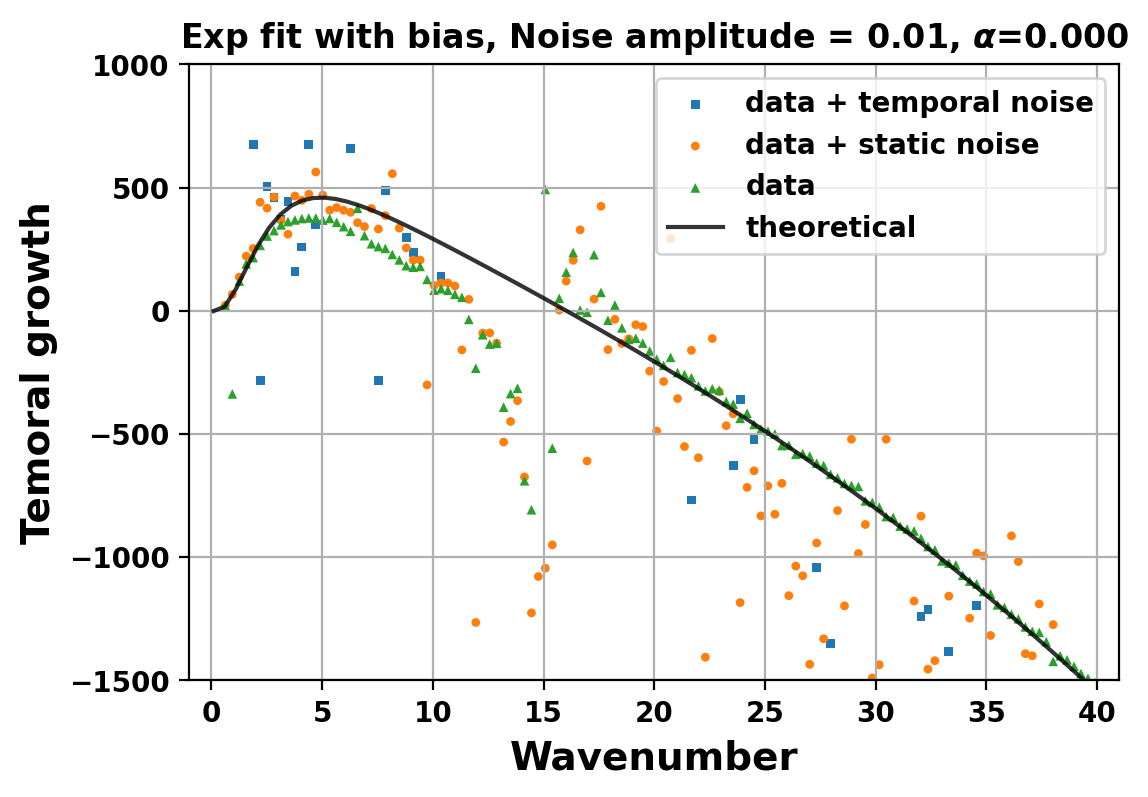}\hfill
    \includegraphics[width=0.49\linewidth]{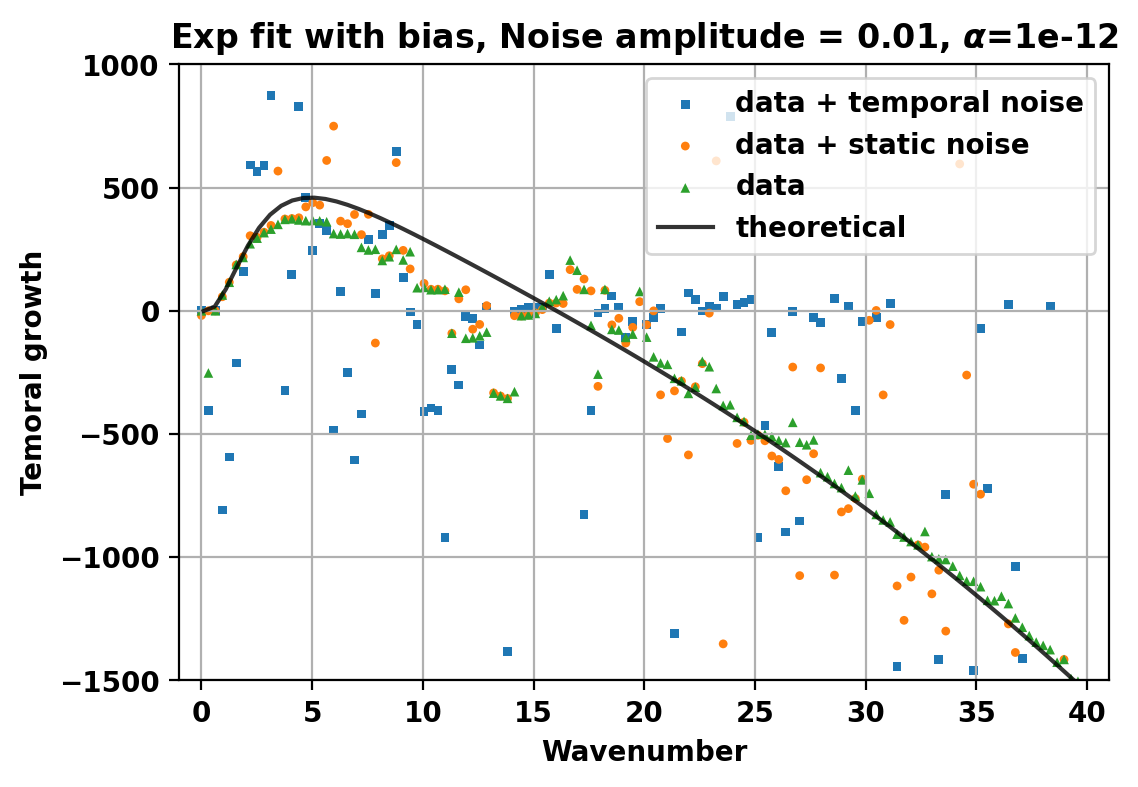}

    \includegraphics[width=0.49\linewidth]{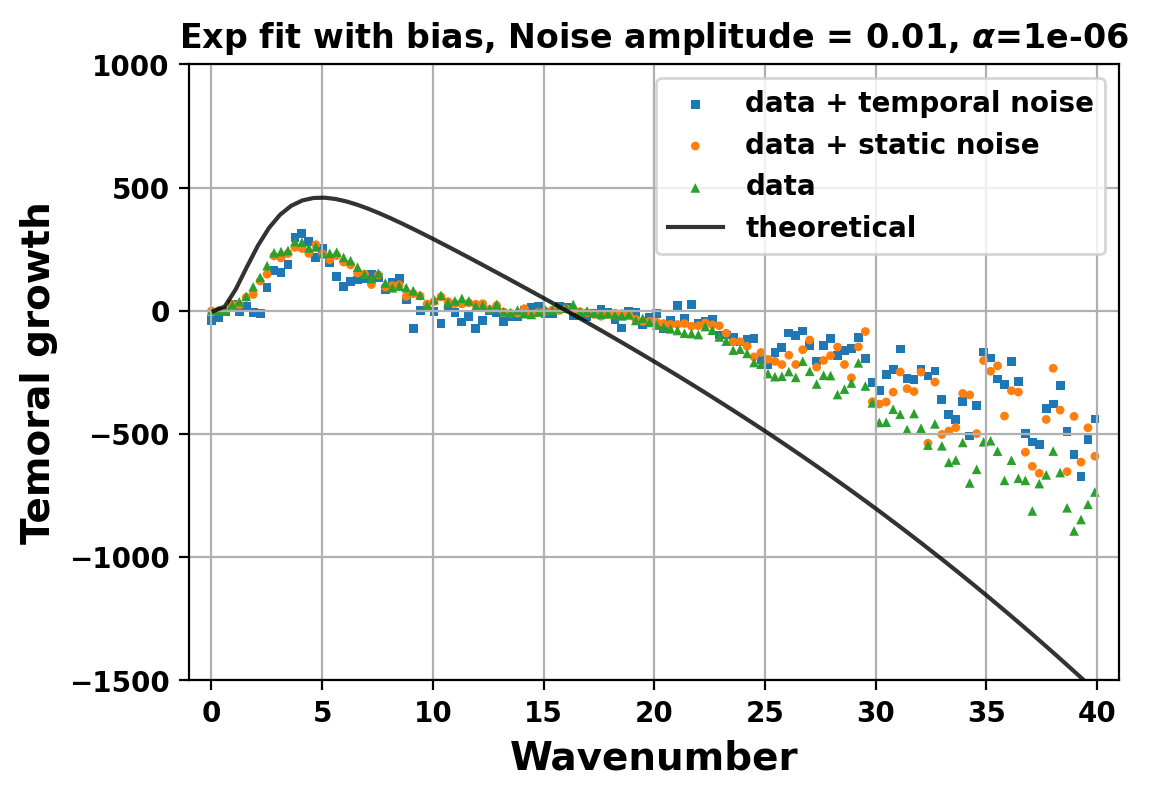}\hfill
    \includegraphics[width=0.49\linewidth]{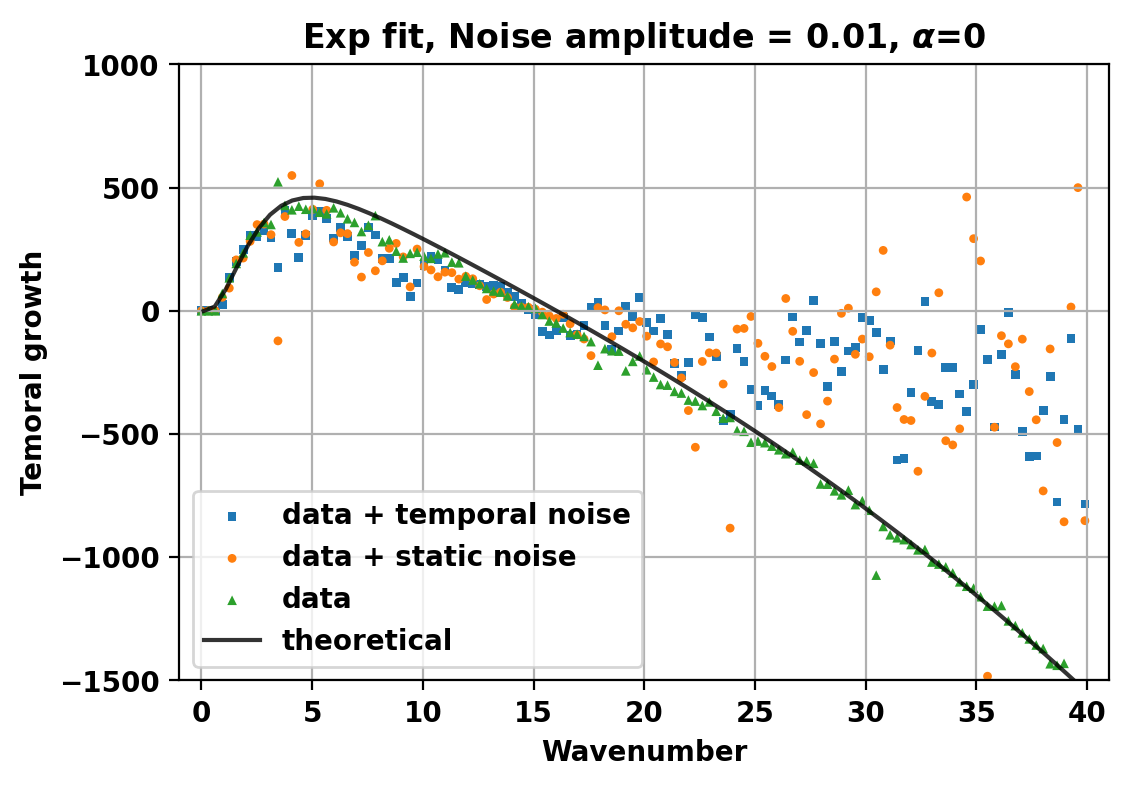}
    \caption{Growth increments for $Ra_m=200$ and $Ra_g=0$. The black curve shows the analytic values, the numeric increments are calculated using (\ref{eq:fft_growth_increments_from_ratio}). The green triangles show the values from noiseless data, the blue squares show the values from data with time-varying noise and the orange circles show the values from data with static noise. a) method 1 with a noise amplitude of 0.001 b) method 1 with a noise amplitude of 0.01, c) method 2 with a noise amplitude of 0.001, d) method 3 with a noise amplitude of 0.001 and a denoising parameter $\beta=1$, e) method 4 with a noise amplitude of 0.01, f) method 4 with a noise amplitude of 0.01 and a regularisation parameter $\alpha=10^{-12}$, g) method 4 with a noise amplitude of 0.01 and a regularisation parameter $\alpha=10^{-6}$, h) method 5 with a noise amplitude of 0.01.}
    \label{fig:k_lambda_grouped_200_methods_1_to_5}
\end{figure}

The magnetic microconvection was simulated in a $20\times20$ domain with a grid of $512\times512$. The timestep was $dt=10^{-5}$ and in total 200 steps were simulated.
% Figures \ref{fig:k_lambda_numeric_fft_noise_1e-3_200_steps_ram_200}, \ref{fig:k_lambda_numeric_fft_noise_1e-2_200_steps_ram_200} show the results for method 1 for $Ra_m=200$ and $Ra_g=0$.
Fig.\ref{fig:k_lambda_grouped_200_methods_1_to_5}a and Fig.\ref{fig:k_lambda_grouped_200_methods_1_to_5}b show the results for method 1 for $Ra_m=200$ and $Ra_g=0$.
The Black curve represents the analytic values (see appendix \ref{appendix:analytic_increments}), the green squares represent the calculated growth increments for data without added noise, the blue squares represent the growth increments for data with added time-dependent noise and the orange circles represent the growth increments for data with added time-independent noise. For the noiseless data this method gives good results for most spatial wavenumbers except for wavenumbers close to zero where there is a bias towards zero. It is expected that the measured values will be lower then the ones predicted by the analytic model because the analytic model assumes a perfectly sharp initial boundary between both fluids which is not the case in the simulations. The slightly smeared-out boundary leads to a smaller gradient in the magnetic potential in (\ref{eq:flow_eqn_dimensionless}) which reduces the magnetic force and the growth increments. If the amplitude of the noise is on the order of $10^{-2}$ or higher then the calculated growth increments greatly differ from the expected values and no clear trends are noticeable. This is unacceptable considering that experimentally captured image data will contain some amount of noise which will destroy the measurement.

% To remedy this method 2 includes a denoising step on the 2D time-series data. Figure \ref{fig:k_lambda_numeric_fft_denoised_noise_1e-2_200_steps_ram_200} shows the results for method 2.
To remedy this method 2 includes a denoising step on the 2D time-series data. Fig.\ref{fig:k_lambda_grouped_200_methods_1_to_5}c shows the results for method 2.
There is an improvement in the results because the noisy points are on average more closely scattered around the expected values, however the deviations are still quite large.

Method 3 is similar to method 2, however here the ratios in (\ref{eq:fft_growth_increments_from_ratio}) are taken from the numeric differences of the Fourier coefficients between consecutive steps in time. This is done to attempt to mitigate any time-independent bias in the input data.
% Unfortunately the results are very inaccurate as seen in figure \ref{fig:k_lambda_numeric_fft_derivative_denoised_noise_1e-2_200_steps_ram_200}.
Unfortunately the results are very inaccurate as seen in Fig. \ref{fig:k_lambda_grouped_200_methods_1_to_5}d.
The resulting growth increments are very sensitive to noise and a lot of the values are scattered outside of the bounds of the figure. There also is a striking deviation from the theoretical curve near $k=15$. Near these wavenumbers the growth increment is expected to be near zero therefore it is possible that the deviations are from second-order effects that are stronger then the first order effects used in the analytic model.

The next approaches use exponential curve fitting instead of ratios. Method 4 fits an exponent plus a constant to each spatial Fourier component over time.
% Figures \ref{fig:k_lambda_numeric_exp_with_c_noise_1e-2_alpha_0_200_steps_ram_200}, \ref{fig:k_lambda_numeric_exp_with_c_noise_1e-2_alpha_1e-12_200_steps_ram_200}, \ref{fig:k_lambda_numeric_exp_with_c_noise_1e-2_alpha_1e-6_200_steps_ram_200} show the results for method 4 for different values of the regularization parameter $\alpha$ from (\ref{eq:fit_penalty}).

Fig.\ref{fig:k_lambda_grouped_200_methods_1_to_5}e, Fig.\ref{fig:k_lambda_grouped_200_methods_1_to_5}f, Fig.\ref{fig:k_lambda_grouped_200_methods_1_to_5}g show the results for method 4 for different values of the regularization parameter $\alpha$ from (\ref{eq:fit_penalty}).
% One can notice that the growth increments for spatial wavenumbers near zero closely follow the analytic curve in contrast to the values obtained by the previous methods.
% Figure \ref{fig:k_lambda_numeric_exp_with_c_noise_1e-2_alpha_0_200_steps_ram_200} shows the fitted values for the case without any regularisation. Interestingly the calculated values for the noiseless data behave unexpectedly around the spatial wave numbers where they are expected to be close to zero.
Fig. \ref{fig:k_lambda_grouped_200_methods_1_to_5}e shows the fitted values for the case without any regularisation. Interestingly the calculated values for the noiseless data behave unexpectedly around the spatial wave numbers where they are expected to be close to zero.
One likely explanation is that her the fit is poor because second-order effects are stronger than the first-order effects that are captured in the analytic model. This erratic behaviour can be suppressed by penalizing large values in the exponent in cases where the improvement in the fit is low by increasing the regularisation parameter $\alpha$. Fig. \ref{fig:k_lambda_grouped_200_methods_1_to_5}f, \ref{fig:k_lambda_grouped_200_methods_1_to_5}g show how different values of $\alpha$ influence the fitted values. A small value of $\alpha$ reduces the erratic values near $k=15$, while a much larger values of $\alpha$ makes the data less prone to noise but also significantly biases the growth increments towards zero. Ideally $\alpha$ should be chosen small enough to only influence the parts where the initial fit is poor.

One might expect that fitting an exponent with a bias will give the most accurate fit, but it turns out that the added bias makes the regression an ill-posed problem where there are many local minima in the penalty function (\ref{eq:fit_penalty}) and for certain cases the location of the minima can be very sensitive to noise.
% This is reflected in Figure \ref{fig:k_lambda_numeric_exp_with_c_noise_1e-2_alpha_0_200_steps_ram_200} where adding time-dependent noise distorts the values so much that most of them are out of the bounds of the plot.
This is reflected in Fig.\ref{fig:k_lambda_grouped_200_methods_1_to_5}e where adding time-dependent noise distorts the values so much that most of them are out of the bounds of the plot.

\begin{figure}[H]
    \includegraphics[width=0.49\linewidth]{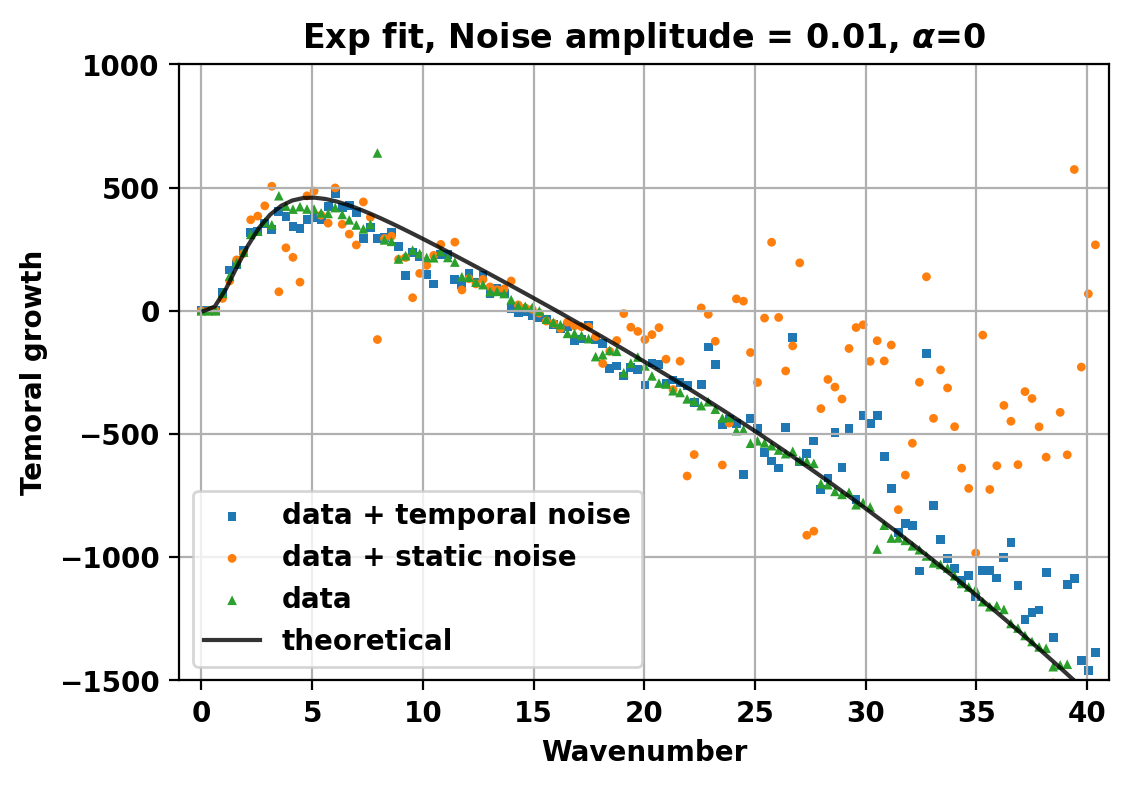}\hfill
    \includegraphics[width=0.49\linewidth]{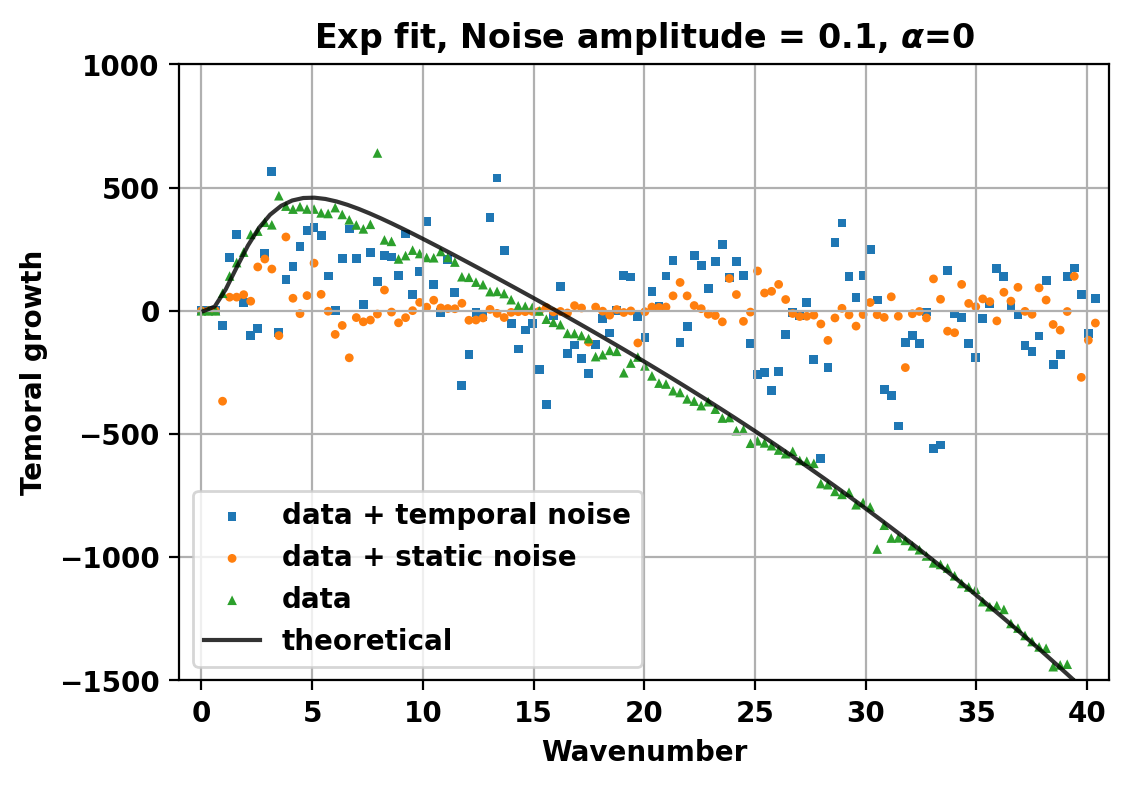}

    \includegraphics[width=0.49\linewidth]{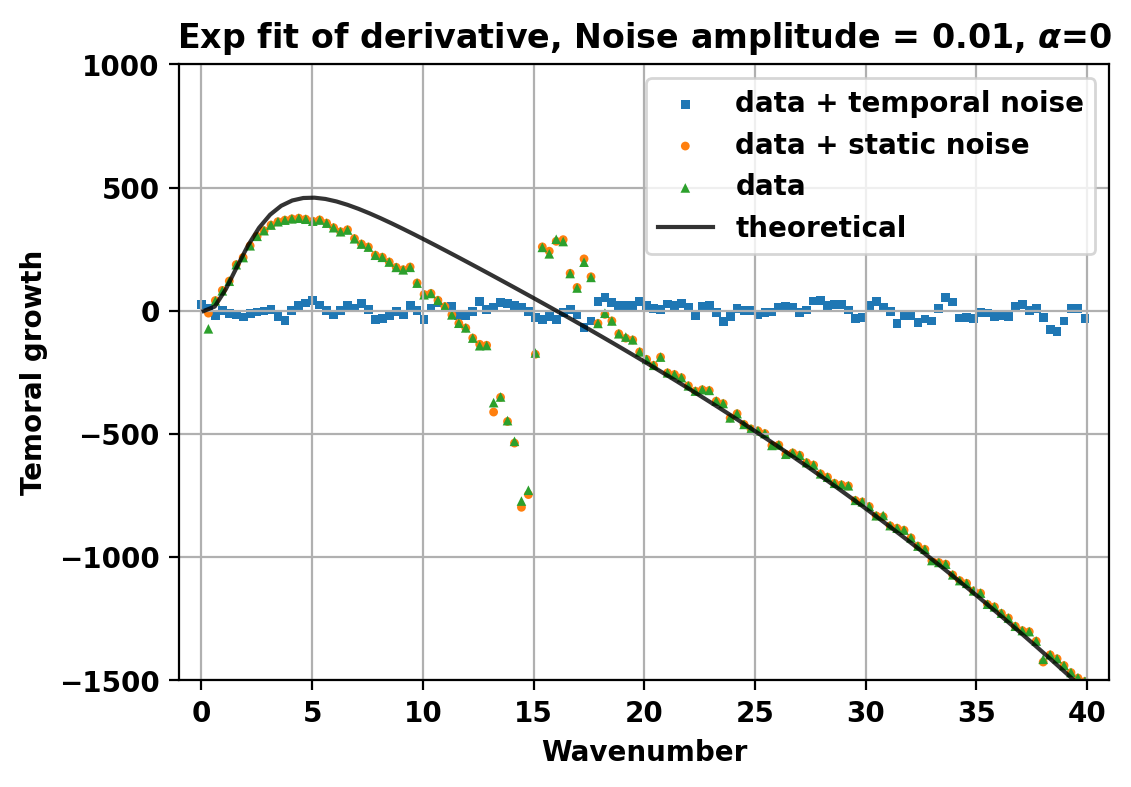}\hfill
    \includegraphics[width=0.49\linewidth]{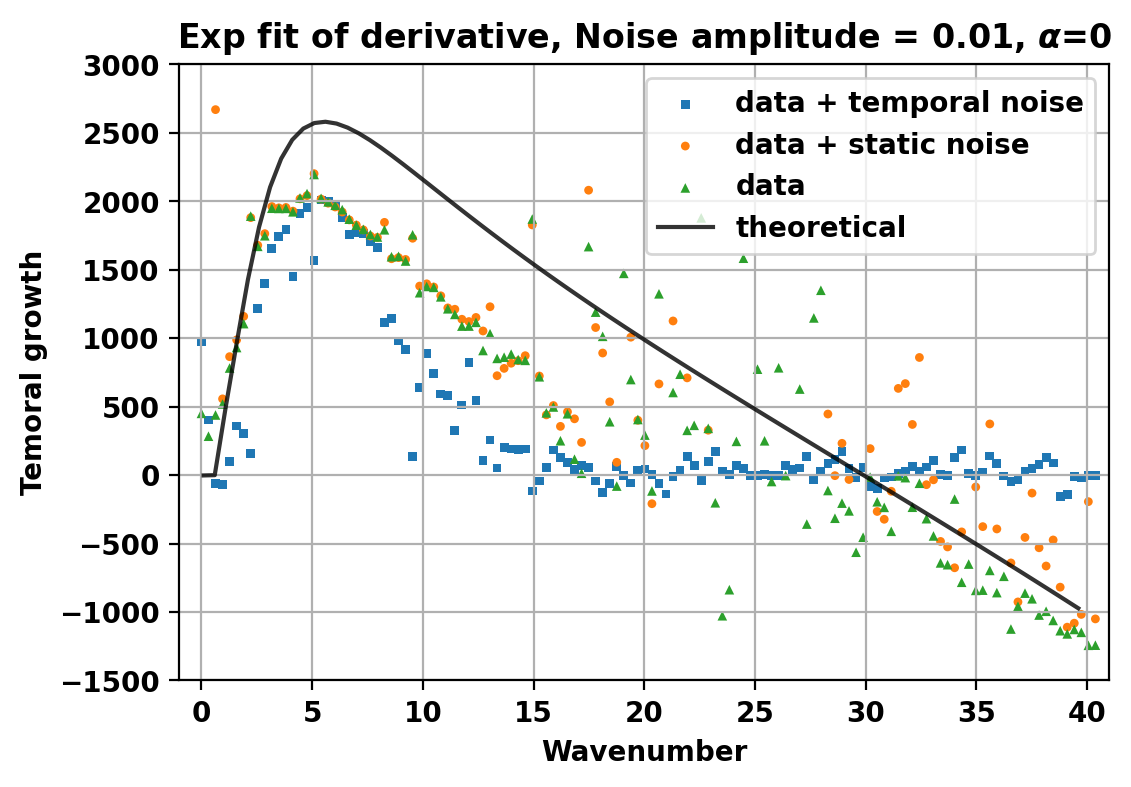}
    
    \caption{Calculated growth increments for various cases. The black curve shows the analytic values, the numeric increments are calculated using (\ref{eq:fft_growth_increments_from_ratio}). The green triangles show the values from noiseless data, the blue squares show the values from data with time-varying noise and the orange circles show the values from data with static noise. a) method 6 for $Ra_m=200$ and $Ra_g=0$ with a noise amplitude of 0.01 and a denoising parameter $\beta=1$, b) method 6 for $Ra_m=200$ and $Ra_g=0$ with a noise amplitude of 0.1 and a denoising parameter $\beta=1$, c) method 7 for $Ra_m=200$ and $Ra_g=0$ with a noise amplitude of 0.01, d) method 8 for $Ra_m=1000$ and $Ra_g=0$ with a noise amplitude of 0.01 and a denoising parameter $\beta=1$.}
    \label{fig:k_lambda_grouped_200_methods_6_to_8}
\end{figure}

Fitting an exponent without a bias is a lot less problematic therefore it is worth trying even when there is a bias caused by time-independent noise.
% Figure \ref{fig:k_lambda_numeric_exp_no_c_noise_1e-2_200_steps_ram_200} shows the growth increments that are calculated by fitting an exponent without a bias according to Method 5. 
Fig.\ref{fig:k_lambda_grouped_200_methods_1_to_5}h shows the growth increments that are calculated by fitting an exponent without a bias according to Method 5. 
% Here the results are less sensitive to noise, and by adding a denoising step as in Method 6 (see fig. \ref{fig:k_lambda_numeric_exp_no_c_denoised_noise_1e-1_200_steps_ram_200}) the results become slightly better.
Here the results are less sensitive to noise, and by adding a denoising step as in Method 6 (see Fig. \ref{fig:k_lambda_grouped_200_methods_6_to_8}a) the results become slightly better.

% Method 7 is similar to method 6 but here the exponent is fitted to the time derivative of the Fourier coefficients. This is done in Figure \ref{fig:k_lambda_numeric_exp_derivative_noise_1e-2_alpha_0_200_steps_ram_200}.
Method 7 is similar to method 6 but here the exponent is fitted to the time derivative of the Fourier coefficients. This is done in Fig.\ref{fig:k_lambda_grouped_200_methods_6_to_8}c.
Here the static noise has no effect, however any temporal noise is gets amplified and completely overpowers any data.
% Another problem is the deviation from the true values near $k=15$ similar to the deviations seen in figure \ref{fig:k_lambda_numeric_fft_derivative_denoised_noise_1e-2_200_steps_ram_200} which uses method 4.
Another problem is the deviation from the true values near $k=15$ similar to the deviations seen in Fig.\ref{fig:k_lambda_grouped_200_methods_1_to_5}d which uses method 4.
Method 8 adds denoising to the concentration field before extracting the Fourier coefficients.
% This is done in figure \ref{fig:k_lambda_numeric_exp_derivative_denoised_noise_1e-3_alpha_0_200_steps_ram_200} and \ref{fig:k_lambda_numeric_exp_derivative_denoised_noise_1e-2_alpha_0_200_steps_ram_200} where different levels of noise were examined.
This is done in Fig.\ref{fig:combined_figure_200_1e-3_200_1e-2_1000_1e-2_1000_1e-3}a and Fig.\ref{fig:combined_figure_200_1e-3_200_1e-2_1000_1e-2_1000_1e-3}b where different levels of noise were examined.
If the noise is very weak then the calculated growth increments approximately follow the shape of the theoretical curve however if the noise is too strong then the calculated values drastically deviate from the theoretical values.
% Figure \ref{fig:k_lambda_numeric_exp_derivative_denoised_noise_1e-2_alpha_0_200_steps_ram_200} also shows that the denoising can potentially negatively affect the calculated values even when there is noise that is constant in time
% Figure \ref{fig:k_lambda_numeric_exp_derivative_denoised_noise_1e-2_alpha_0_200_steps_ram_200} also shows that the denoising can potentially negatively affect the calculated values even when there is noise that is constant in time.
Fig.\ref{fig:combined_figure_200_1e-3_200_1e-2_1000_1e-2_1000_1e-3}b also shows that the denoising can potentially negatively affect the calculated values even when there is noise that is constant in time.
Despite the many flaws, Method 8 has the very desirable property of being able to ignore additive bias in the data which is very relevant for examining data from real experiments. The main source for temporal noise is camera noise however the main sources of static noise are biases in the concentration field that are caused by specks and smudges on the Hele-Schaw cell and also uneven lighting.

% For comparison method 8 was also used on simulated data for $Ra_m=1000$ and the results are shown in figures \ref{fig:k_lambda_numeric_exp_derivative_denoised_noise_1e-3_alpha_0_200_steps_ram_1000}, \ref{fig:k_lambda_numeric_exp_derivative_denoised_noise_1e-2_alpha_0_200_steps_ram_1000}, and \ref{fig:k_lambda_numeric_exp_derivative_denoised_noise_1e-2_alpha_1e-12_200_steps_ram_1000}.
% For comparison method 8 was also used on simulated data for $Ra_m=1000$ and the results are shown in figures \ref{fig:combined_figure_200_1e-3_200_1e-2_1000_1e-2_1000_1e-3}c, \ref{fig:k_lambda_numeric_exp_derivative_denoised_noise_1e-2_alpha_0_200_steps_ram_1000}, and \ref{fig:combined_figure_200_1e-3_200_1e-2_1000_1e-2_1000_1e-3}d.
For comparison method 8 was also used on simulated data for $Ra_m=1000$ and the results are shown in Fig.\ref{fig:combined_figure_200_1e-3_200_1e-2_1000_1e-2_1000_1e-3}c, Fig.\ref{fig:k_lambda_grouped_200_methods_6_to_8}d, and Fig.\ref{fig:combined_figure_200_1e-3_200_1e-2_1000_1e-2_1000_1e-3}d.
Since the perturbation growth is more intense the perturbations are more distinguishable from noise therefore the noise has a smaller effect on the measured values.
% In figures \ref{fig:k_lambda_numeric_exp_derivative_denoised_noise_1e-2_alpha_0_200_steps_ram_1000}, and \ref{fig:k_lambda_numeric_exp_derivative_denoised_noise_1e-2_alpha_1e-12_200_steps_ram_1000}
% In figures \ref{fig:k_lambda_numeric_exp_derivative_denoised_noise_1e-2_alpha_0_200_steps_ram_1000}, and \ref{fig:combined_figure_200_1e-3_200_1e-2_1000_1e-2_1000_1e-3}d it is apparent that the largest growth increments are largely unaffected by noise however smaller nonzero values are dragged towards zero when there is time-dependent noise.
In Fig.\ref{fig:k_lambda_grouped_200_methods_6_to_8}d, and Fig.\ref{fig:combined_figure_200_1e-3_200_1e-2_1000_1e-2_1000_1e-3}d it is apparent that the largest growth increments are largely unaffected by noise however smaller nonzero values are dragged towards zero when there is time-dependent noise.
This fact is important to keep in mind when looking at measurements from noisy data.


\newpage

\section{Image data preprocessing.}
\label{appendix:image_despeckling}

It is often the case that the experimental Hele-Schaw cell contains some amount of impurities either from the magnetic fluid itself or the construction of the cell. These impurities present themselves as many static darkened specks in the image and they have a risk of adding errors to the growth increment measurements. These specks were removed using the following steps:

\begin{algorithm}%[H]
    \caption{Image speck detection and removal}
    Let $I(x,y)$ be the input image for the measured light intensity.
    \begin{enumerate}
        \item Apply Gaussian smoothing with $\sigma_{c}$ to $I$ to get $I_{c}$.
        \item Apply a difference of Gaussians with $\sigma_1$ and $\sigma_2$ to $I$ to get $I_{f}$.
        \item Compute image curvature of $I_{c}$ using
        \begin{equation}
            k_{c} = \nabla\cdot\qty(\frac{\nabla I_{c}}{\sqrt{\qty|\nabla I_{c}|^2+\epsilon}})
        \end{equation}
        where $\epsilon_{c}$ is a regularization parameter to filter out small gradients.
        \item Compute the coarse speck indicator function
        $m'_{c}$
        \begin{equation}
            m'_{c} = \qty(k_{c}+a\nabla I_{c})
        \end{equation}
        where $a$ is a scaling coefficient.
        \item Compute the fine speck indicator function
        $m'_{f}$ by taking the absolute difference between the maximum and minimum value of pixels in a circular neighbourhood around each pixel with a integer radius of $n_f$.
        % curvature_fine = skimage.filters.rank.gradient(img_smoothed_fine, footprint=skimage.morphology.disk(4))
        \item Compute the coarse speck mask
        $m_{c}$
        \begin{equation}
            m_{c} = 
            \begin{cases}
                1,& \text{if } m'_{c} \geq p_c\\
                0,              & \text{otherwise}
            \end{cases}
        \end{equation}
        where $p_c$ is the n-th percentile of the values of $m'_{c}$.
        \item Compute the fine speck mask $m_{fine}$
        \begin{equation}
            m_{f} = 
            \begin{cases}
                1,& \text{if } m'_{f} \geq p_f\\
                0,              & \text{otherwise}
            \end{cases}
        \end{equation}
        where $p_f$ is the n-th percentile of the values of $m'_{f}$.
        \item Compute the final mask $m_{final}$ by taking the maximum of $m_{c}$ and $m_{f}$ at each point.
        \item Preform morphological dilation of $m_{final}$ with a disk-shaped kernel of radius $n_d$
        \item Preform morphological area-closing \cite{vincentGrayscaleAreaOpenings1993} of $m_{final}$ where segments larger than $N$ pixels are set to zero.
        \item (Optional) set all boundary values of $m_{final}$ to zero. This can prevent inpainting artefacts for specks near the boundaries.
        \item Inpaint the image using biharmonic inpainting \cite{damelinSurfaceCompletionImage2018} using $m_{final}$ as the mask.
    \end{enumerate}
    
    \label{alg:despeckling}
\end{algorithm}

The variables $\sigma_c$, $\sigma_1$, $\sigma_2$, $\epsilon$, $a$, $n$, $p_c$, $p_f$ are scalar variables that can be adjusted in order to provide the desired results. In our case the the values that were chosen were $\sigma_c=20$, $\sigma_1=2$, $\sigma_2=3$, $\epsilon=0.1$, $a=10$, $p_c=85\%$, $p_f=95\%$. This filter is designed to remove small round specks and medium sized smudges while ignoring the gradients of the fluid interface. The curvature of the blurred image is a good indicator for blurry smudges and the difference of Gaussians followed by local differences is a good indicator of small dot shaped specks. Both the coarse and fine speck detection steps ignore slow line-shaped gradients.

Since the speckles are just extra objects that absorb light according to the Lambert law it is save to assume that their effect on the image is multiplicative. These specks are static therefore it is possible to despeckle an entire image sequence by multiplying each image with the ratio of one selected despeckled frame and its original copy. Since our despeckling algorithm works best if the interface has zero curvature the initial frame where the interface was flat was chosen for despeckling.

\end{document}